# Ultra-long-range spin coupling in graphene revealed by atomically resolved spin excitations

**Authors:** B. Viña-Bausá[1]*, A. T. Costa[2,3], J. Henriques[2,4], E. Cortés-del Río[5], R. Carrasco[1], P. Mallet[6], J-Y. Veuillen[6], J. Fernández-Rossier[2]*, I. Brihuega[1,7]*

**Affiliations:**

[1]*Departamento de Física de la Materia Condensada, Universidad Autónoma de Madrid; E-28049 Madrid, Spain*

[2]*International Iberian Nanotechnology Laboratory (INL); Avenida Mestre José Veiga, 4715-310 Braga, Portugal*

[3]*Physics Center of Minho and Porto Universities (CF-UM-UP), Universidade do Minho; Campus de Gualtar, 4710-057 Braga, Portugal*

[4]*Universidade de Santiago de Compostela, Santiago de Compostela, Spain.*

[5]*Department of Physics, University of Hamburg; D-20355 Hamburg, Germany.*

[6]*Université Grenoble Alpes, Grenoble, F-38400 France and CNRS, Institut Néel; Grenoble, F-38042 France*

[7]*Condensed Matter Physics Center (IFIMAC) and Instituto Nicolás Cabrera (INC), Universidad Autónoma de Madrid; E-28049 Madrid, Spain*

* Corresponding authors: beatriz.vina@uam.es   ivan.brihuega@uam.es

Magnetic interactions between localized spins-½ play a central role in quantum magnetism, spin-based quantum computing, and quantum simulation. The range and strength of these interactions are key figures of merit. Here, we probe exchange interactions in pairs and trimers of spins-½ introduced by chemisorption of individual hydrogen atoms on graphene. Using scanning tunneling microscopy and inelastic electron tunneling spectroscopy, supported by large-scale mean-field Hubbard calculations, we demonstrate 3 meV exchange couplings at separations beyond 10 nm, surpassing all prior systems. The couplings can be ferro- or antiferromagnetic depending on the relative sublattice arrangement. Real-space mapping of spin excitation amplitudes enables characterization with atomic-resolution. Through atomic manipulation we extend this control to spin trimers, revealing collective spin excitations when pairwise exchange couplings are comparable.

# Introduction

Understanding and controlling magnetic coupling at a fundamental level is crucial for spin-based quantum technologies[1-3]. In this context, research efforts in the last decades have focused on enhancing spin lifetimes, coherence times and, importantly, on pushing the limits of the spatial and energy scales[3-9]. Different platforms have been investigated, including semiconductor quantum dots[10-12]; electronic and nuclear spins of individual dopants in bulk, solid-state systems[13-15] or electron spins provided by single atoms or molecules on surfaces[16,17]. However, it still remains a challenge to combine in the same system atomic scale control with sufficiently strong long-range interactions that provide thermal stability. For instance, for artificial structures made with atomic manipulation of $d$ and $f$ transition metal atoms on surfaces, the range of magnetic coupling is restricted to the atomic scale and exchange energies are of the order of a few meV[18,19]. More recently, carbon-based molecular spin systems obtained via on-surface synthesis have gained attention because of their chemical versatility, weak spin-orbit coupling and larger exchange energies[20–26]. Still, in these nanoscale systems, exchange with the conducting surface is expected to limit the spin lifetimes[19,27].

Here, we study a $S=1/2$ point-defect, atomic hydrogen chemisorbed in graphene[28], whose magnetic moment extends, and strongly interacts, significantly beyond the atomic scale. Atomically modified graphene constitutes a sweet spot between solid-state and atom by atom approaches, where long-range interactions coexist with large exchange energies. Together with this, our system also combines two exceptional peculiarities: we can have coexisting ferromagnetic and antiferromagnetic interactions, and our platform enables carrying out atomic scale manipulations. With all this, we demonstrate that graphene provides a unique playground to understand, explore and control the interactions between $S=1/2$ magnetic moments.

By Scanning Tunneling Microscopy/Spectroscopy (STM/STS), we build and characterize our magnetic configurations: we use the STM tip to selectively introduce $S=1/2$ magnetic moments by attaching single H atoms to graphene. Our atomically resolved inelastic electron tunneling spectroscopy (IETS) data probes the spin excitations of pairs and trios of H atoms on graphene. Combined with large-scale mean-field Hubbard (MFH) and Random Phase Approximation (RPA) calculations, these results provide a systematic study as well as real space visualization and quantification of the exchange interactions as a function of distance and sublattice.

**Inducing $S=1/2$ magnetic moments in graphene with single H atoms**

Our work relies on the fact that an individual H atom covalently attached to graphene leaves an unpaired electron and therefore induces a $S=1/2$ magnetic moment[28-30]. When the C-H bond is formed, both the carbon $p_z$ orbital and one electron from the π cloud are taken away from graphene. As a result, a quasilocalized state emerges at the Dirac point, known as zero mode[29]. Because of Coulomb repulsion, double occupation of this state is blocked, resulting in the formation of a local moment. The density of states (DOS) of a single H atom in graphene is characterized by the presence of two narrow peaks at the Fermi level ($E_F$), whose energy splitting, in the range of 10 meV, is a metric of the double occupation energy overhead, that slightly depends on the local environment[28], on account of the quasi-localized nature of the state[29]. These peaks correspond to the addition and removal of an electron from the quasi-localized state, taking as reference the singly occupied state.

Magnetic moments induced in graphene by H atoms have been observed to extend over several nanometers, presenting a threefold symmetry dictated by graphene crystal structure[28]. In this way, on STM images, single H atoms can be identified because they appear as bright protrusions and are surrounded by a threefold, $\sqrt{3}\times\sqrt{3}$ pattern rotated 30º with respect to the graphene lattice. The corresponding absorption sublattice can be inferred because this threefold modulation associated to the magnetic state points in opposite (parallel) directions for H atoms positioned on opposite (equivalent) sublattices. The magnetic moment is essentially induced only on the graphene sublattice opposite to the H absorption site.

The possibility of carrying out atomic scale manipulations has also been demonstrated[28,31] and constitutes a fundamental pillar for this work. Using the STM, H atoms can be selectively attached and removed in graphene.

## Results and Discussion

**Two interacting S=1/2 spins**

We start by building the simplest system that hosts inelastic spin excitations at zero magnetic field: two interacting S=1/2 magnetic moments. We accomplish this by arranging well isolated pairs of H atoms on graphene, ensuring the absence of other adsorbed H atoms at distances less than 10 nanometers. The spin of the ground state of H dimers is determined by the relative sublattice (A or B) adsorption site of the H atoms[28,30] in agreement with Lieb's theorem[32]. When both H atoms are adsorbed on opposite graphene sublattices (AB pair), the two induced S=1/2 couple antiferromagnetically and have a singlet ground state, with S=0. At short separations ($\leq$ 1.5 nm), the AB pair adopts a closed-shell singlet (S=0) state due to strong orbital hybridization[33]. As the AB distance increases, a smooth closed-to-open shell transition occurs. Conversely, for two H atoms adsorbed on the same graphene sublattice (AA pairs) the exchange coupling is ferromagnetic, and the ground state is an open-shell triplet (S = 1), regardless of separation. Given the very small spin-orbit coupling in graphene, magnetic anisotropy is negligible and such triplet state is degenerate at zero magnetic field[34].

According to the selection rule imposed by the conservation of angular momentum, possible spin excitations should satisfy $\Delta S=0$ or $\Delta S=\pm 1$[18]. For AB pairs, with a singlet ground state (S=0), spin transitions to the triplet excited state (S=1) can be induced by inelastic tunneling electrons, while the AA pair will have a triplet ground state (S=1) with possible transitions to the singlet excited state (S=0). Thus, pairs of chemisorbed hydrogen atoms in graphene permit us to create exchange-coupled S=1/2 pairs with both ferro and antiferromagnetic coupling, and, either S=0 or S=1 ground states.

Our results for an antiferromagnetically coupled AB pair, separated 4.4 nm, are shown in Fig. 1.a-c. The spectra, measured with a superconducting tip and here deconvoluted (see also Table S1), present clear energy-symmetric steps at both bias polarities, indicated with black arrows, with a large increase in the differential conductance. This is the fingerprint[18,23] of spin excitations in inelastic electron tunneling spectroscopy (IETS). The spin excitation takes place at 17 meV and corresponds to the transition from the singlet ground state to the triplet excited state (see methods for details). We have included a scheme of the energy levels to illustrate this. Additionally, Fig. 1c shows a MFH calculation of the atomic magnetization for this precise AB configuration, where red and blue indicate opposite spin directions, confirming the antiferromagnetic ground state of the AB configuration.

The analogous ferromagnetic situation is illustrated in Fig. 1d-f. Here, we arrange an AA pair at 2.4 nm to create a system with a ground state of total spin S=1 (Fig. 1d). Again, we observe inelastic features at 17 mV, indicated with arrows (Fig. 1e), corresponding to the triplet to singlet excitation of such configuration and a ferromagnetic ground state in the MFH magnetization calculation (Fig. 1f).

In the AA case, in addition to the symmetric inelastic transitions, we systematically observe spectral features at low energy, similar to those of the monomers. In this regard, AA pairs are a unique system, where the addition energies are smaller than the inelastic spin excitation energies. Beyond this addition-energy voltage, the occupation of the hydrogen-induced zero modes is fluctuating between 1 and either 0 (negative bias) or 2 (positive bias). Therefore, at that bias, there is still a finite probability that AA pair behaves like a S=1/2 dimer. Additional data of H pairs and their $dI/dV$ spectra can be found on the extensive dataset presented in Tables 1 and 2 of the Supplementary Information.

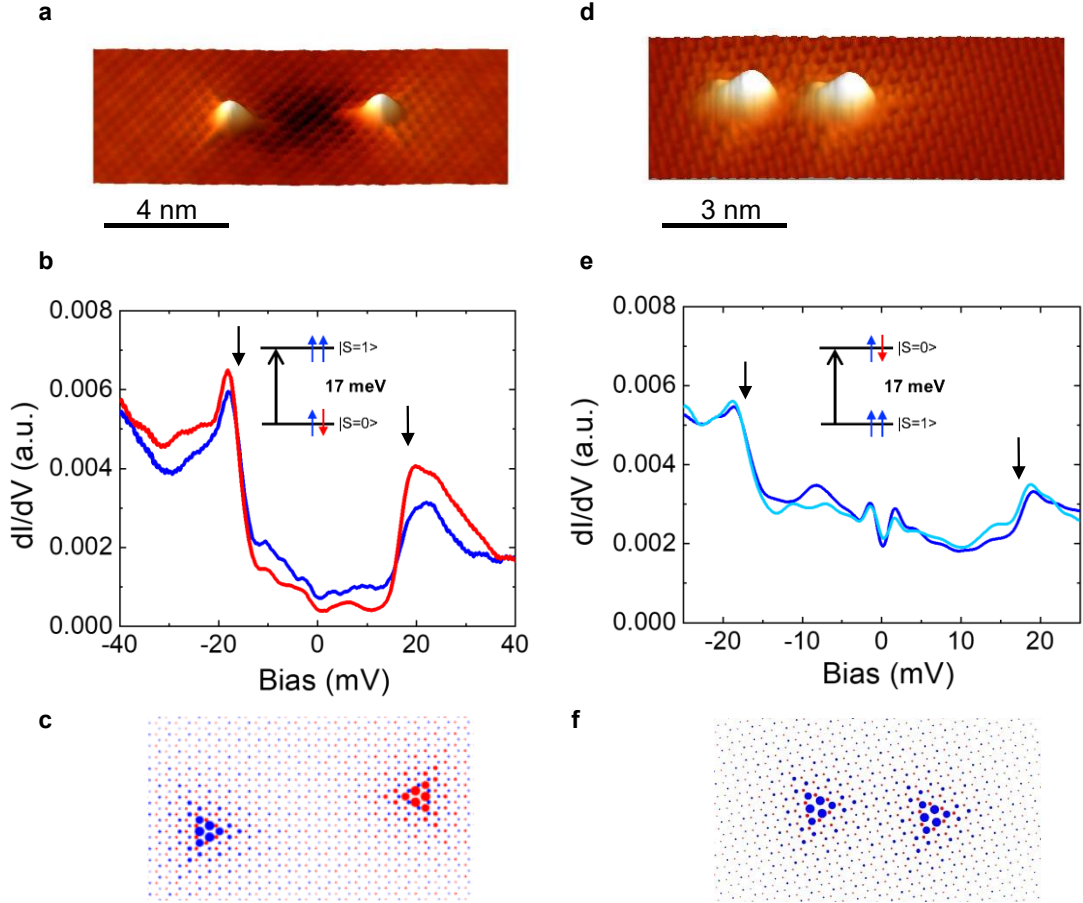

**Fig. 1. Two interacting S=1/2 spins. a,** Topography image of 2H atoms absorbed on opposite graphene sublattices (AB pair) at 4.4 nm (4mV, 0.05nA). **b,** *dI/dV* spectra measured on each of the H atoms (50mV, 0.1nA). Arrows indicate the inelastic steps that correspond to the spin excitations. The singlet to triplet transition energy is 17 meV. **c,** MFH atomic magnetization of the AB configuration. **d,** Topography image of 2H atoms absorbed on the same graphene sublattices (AA pair) at 2.4 nm (30mV, 0.1nA). **e,** *dI/dV* spectra measured on each of the H atoms (30mV, 0.1nA). Arrows indicate the inelastic steps that correspond to the spin excitations. The triplet to singlet transition energy is 17 meV. **f,** MFH atomic magnetization of the AA configuration.

## *Control over the magnetic coupling*

We have systematically arranged many different pairs of H atoms on graphene, both of AA and AB type, performed IETS measurements and extracted the spin excitation energies (see Supplementary Tables 1-2 and Supplementary Fig. 1-2). In Fig. 2a we plot the experimental excitation energy as a function of the distance between H atoms for different configurations. The measured values are compared with MFH calculations (see Fig. 2b) which show the energy difference between the excited and ground state for both AA and AB pairs in different relative positions lying on graphene crystallographic directions. The calculations account for the three main experimental observations: first, interactions observable with IETS, i.e., with more than 3 meV, are observed at separations of up to 10 nm.

Second, for the same distance, coupling in AA pairs is most often weaker than in AB ones. Third, the strength of the exchange coupling does not only depend on distance, but also on the relative orientation of the anisotropically induced magnetic moments. Therefore, the controlled chemisorption of atomic hydrogen on graphene permits to create S=1/2 pairs with three different knobs to tune the strength and sign of their exchange interactions: distance, sublattice and orientation relative to the crystallographic axis.

Several exchange mechanisms are at play. First, direct exchange, relevant for AA pairs, for which the zero-modes overlap in real space, which promotes ferromagnetic exchange. Second, kinetic exchange, relevant for AB pairs, that promotes antiferromagnetic exchange. Third, indirect exchange (RKKY-like) interactions, both for AA and AB cases, mediated by the electrons in the extended states that couple with the local moments induced by the chemisorbed hydrogens. Since local moments are induced in one sublattice only, and the sign of the RKKY interaction in graphene is FM (AF) for AA (AB) pairs, RKKY coupling also contributes to establish the interplay between the sign of the interaction and the sublattice degree of freedom[35].

The range and strength of spin interactions between H induced moments is dramatically larger than in any other platform, as shown in Fig. 2a. where we have included the literature values for exchange couplings in different paired atomic systems (see Supplementary Fig. 3 for more details).

Our length scales are one order of magnitude larger than those required in d and f shells coupled by direct exchange[18,36–39] and, in diradical nanographenes, with exchange interactions in the order of 20 meV, the separation of local moments is at least 5 times smaller than in our case[23,24]. Moreover, the only anticipated competing case with exchange interactions between the electronic spins at comparable long distances are two donors in Silicon, relevant for quantum computing, with theoretical predicted values[13,40] in the range of µeV at a distance of 9 nm, to be compared with the 5 meV in our case. We also note that hypothetical metamaterials built with an artificial lattice of chemisorbed H atoms in graphene[41] would have exchange energies of the order of 20 meV, so that collective magnetic states would be robust at room temperature.

All these extensive experiments and calculations presented in Fig. 2 provide a complete picture of the energy scale of interactions between two $S=1/2$ magnetic moments in graphene.

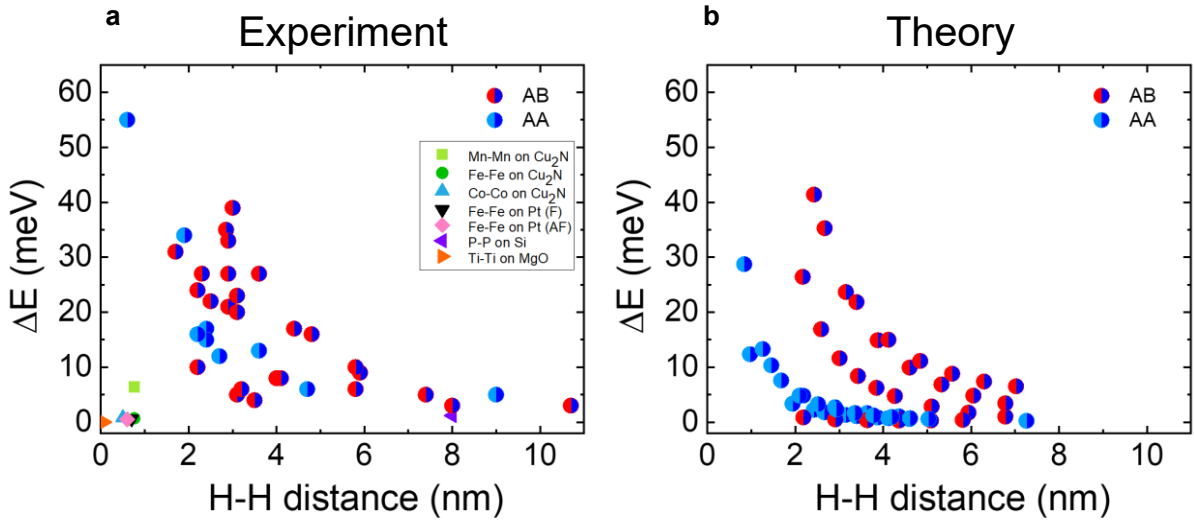

**Fig. 2. Distance dependence of magnetic coupling. a,** Experimental (IETS) excitation energy for AA and AB pairs as a function of the distance between H atoms. Literature exchange values have been included for comparison from refs. *18, 36-40*. **b,** MFH calculation of the excitation energy for AA and AB pairs as a function of the distance between H atoms.

*Spatially mapping spin excitations*

Visualization of *dI/dV* maps at the characteristic spin-excitation energies of selected configurations provides deeper insight into the nature of these coupled states. As shown in Fig. 3, this approach enables spatially resolved mapping of spin excitations, directly probing the spatial extent and coherence of the magnetically coupled states.

In the first example, (Fig. 3a-d) we present an AB pair, placed at 7.4 nm distance, antiferromagnetically coupled with a measured excitation of ΔE=5 meV. As anticipated by the unusually long-range coupling, the *dI/dV* map at this excitation energy shows a prominently bright region, surrounding the H atoms, indicating that we are able to induce the singlet to triplet transition even several nm away from the absorption sites, unveiling the non-local nature of the induced magnetic moment. The amplitude of the *dI/dV* map relates to the spin spectral weight[26]. Our RPA calculations (see Supplementary Fig. 4), show that spin spectral weight map is very similar to the atomic magnetization map obtained in the mean field approximation. Therefore, we can also interpret the measured *dI/dV* maps (Fig. 3c) as a proxy for the local magnetization and compare them with MFH calculated magnetization (Fig. 3d).

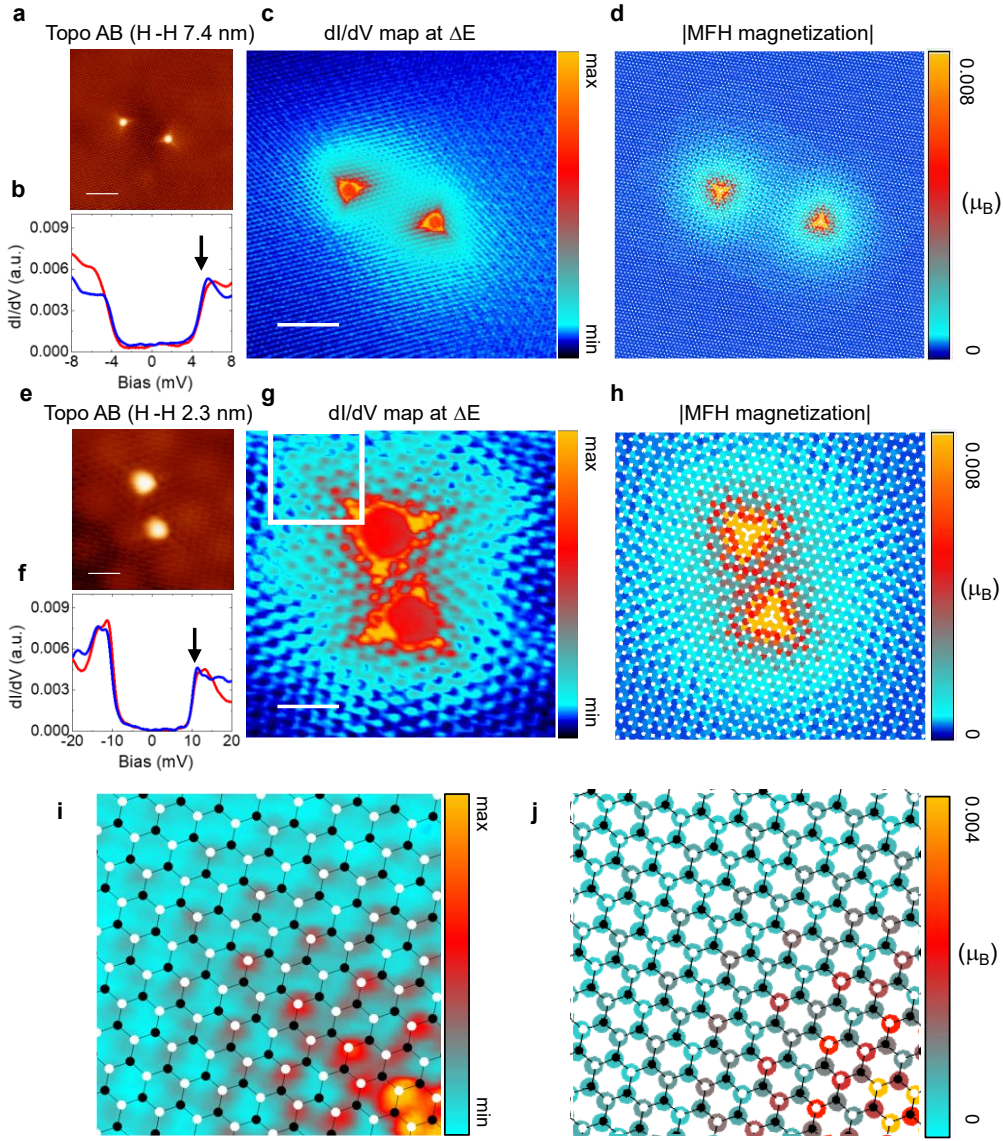

**Fig. 3. Spatially mapping spin excitations. a,** Topography image of an AB pair of H atoms at 7.4 nm distance (6mV, 0.1nA, scalebar: 5 nm). **b,** *dI/dV* spectra measured on each of the H atoms in **a** (12mV, 0.1nA). **c,** *dI/dV* map at the excitation energy (6.3mV, 0.1nA, $V_{rms}$=0.5mV, scalebar: 5 nm). **d,** Magnetization modulus of the antiferromagnetic ground state corresponding to experimental configuration shown in **a** calculated by MFH. **e,** Topography image of an AB pair of H atoms at 2.2 nm distance (98mV, 0.07nA, scalebar: 1.5nm). **f,** *dI/dV* spectra measured on each of the H atoms in **e** (30mV, 0.1nA, scalebar: 1.5nm). **g,** *dI/dV* map at the spin excitation

energy (12mV). **h,** Magnetization modulus of the antiferromagnetic ground state corresponding to the experimental configuration shown in **f** calculated by MFH. **(i)** Zoom of the *dI/dV* map in **g** (white inset) with a superimposed graphene lattice showing sublattice polarization. **j,** Zoom (2.3x2.3 nm$^2$) of the magnetization in **h** a superimposed graphene lattice showing sublattice polarization.

In a second example, (Fig. 3f-Jj) we show an AB pair at a closer distance (2.2 nm), that presents a spin excitation at ΔE=10 meV. The conductance map performed at this energy is again very similar to the MFH local magnetization. The detailed analysis of the interference pattern created on graphene, in both theory and experiment, shows that magnetization, and thus spin excitations, are sublattice polarized around each H atom (see Fig.3 i,j). In other words, placing the tip on either sublattice lets us induce spin transitions with true atomic-scale selectivity. Importantly, while coupling takes place at unusually long distances, we are still able to resolve spin excitations and therefore the magnetization at the atomic scale.

**Spin S=1/2 trimers**

Building on our understanding of the magnetic coupling in 2H atom configurations and exploiting ultralong-range interactions together with STM's atomic-manipulation capabilities, we can now scale up this approach by using individual H atoms as S=1/2 building blocks to assemble larger spin networks. In this next step, we investigate the coupling that emerges when 3H atoms chemisorb on graphene. The relevant space of states for S=1/2 trimers is composed of a quartet (S=3/2) and two doublets (S=1/2) (see SI7). Based on the sublattice of the chemisorption site, trimers come in two flavors, AAA and ABB. The spin of their ground state is expected to be S=3/2 and S=1/2, respectively, on account of the interplay between sublattice and exchange sign discussed above, and also by applying Lieb's theorem. Importantly, once the IETS of a given trimer is taken, it is possible to selectively remove one H, resulting in a dimer, whose exchange energy can be measured.

In the present work we focus on ABB trimers with an S = 1/2 ground state to investigate how distinct coupling regimes, set by the ratios of pairwise spin interactions, are reflected in their IETS spectra (for completeness, Supplementary Fig. 5 presents data for the AAA, S = 3/2 system). Figure 4 presents IETS data for three ABB trimers at similar interatomic separations, and how their spectra evolve when one hydrogen atom is removed. In all cases, the trimer spectra (Figs. 4c, 4f, 4i) differ significantly from those of the corresponding dimers (Figs. 4l, 4o, 4r), clearly demonstrating the presence of exchange interactions between the removed hydrogen atom and the remaining pair. Furthermore, the inelastic excitations observed in the remaining dimer spectra reinforce the existence of exchange coupling between these two hydrogens, as previously discussed. Thus, the data consistently indicates a system comprising three exchange-coupled spins.

Depending on the relative strength of the three pairwise exchange interactions in an ABB trimer, two distinct limiting regimes can be identified. In the first regime, characterized by one spin on the B sublattice being weakly coupled and the remaining AB pair strongly coupled (Fig. 4a–c), the IETS spectrum of the weakly coupled spin significantly deviates from that of the strongly coupled dimer. It lacks spin excitations and instead resembles the spectrum of an isolated hydrogen atom. In contrast, the strongly coupled pair displays a spectrum similar to those shown in Figures 1 and 3, featuring a single excitation step. This regime corresponds to bipartite entanglement.

In the second regime, the antiferromagnetic exchange couplings are comparable in magnitude and collective spin excitations dominate. Consequently, we expect similar spectra across all three atomic sites. This behavior is clearly illustrated in Fig. 4g-i, where spectra on all three sites share a common excitation feature at around 34 meV, indicative of tripartite entanglement. Furthermore, ABB trimers provide an excellent framework to investigate intermediate coupling scenarios between these two extremes. Such an intermediate situation is illustrated by the trimer shown in Fig. 4d–f, where a strongly coupled AB dimer exhibits a common excitation at 20 meV (data points 1 and 2), also visible in the spectrum of the third, less coupled B spin (data point 3), albeit with significantly reduced intensity. We classify this scenario as a weak dimer state.

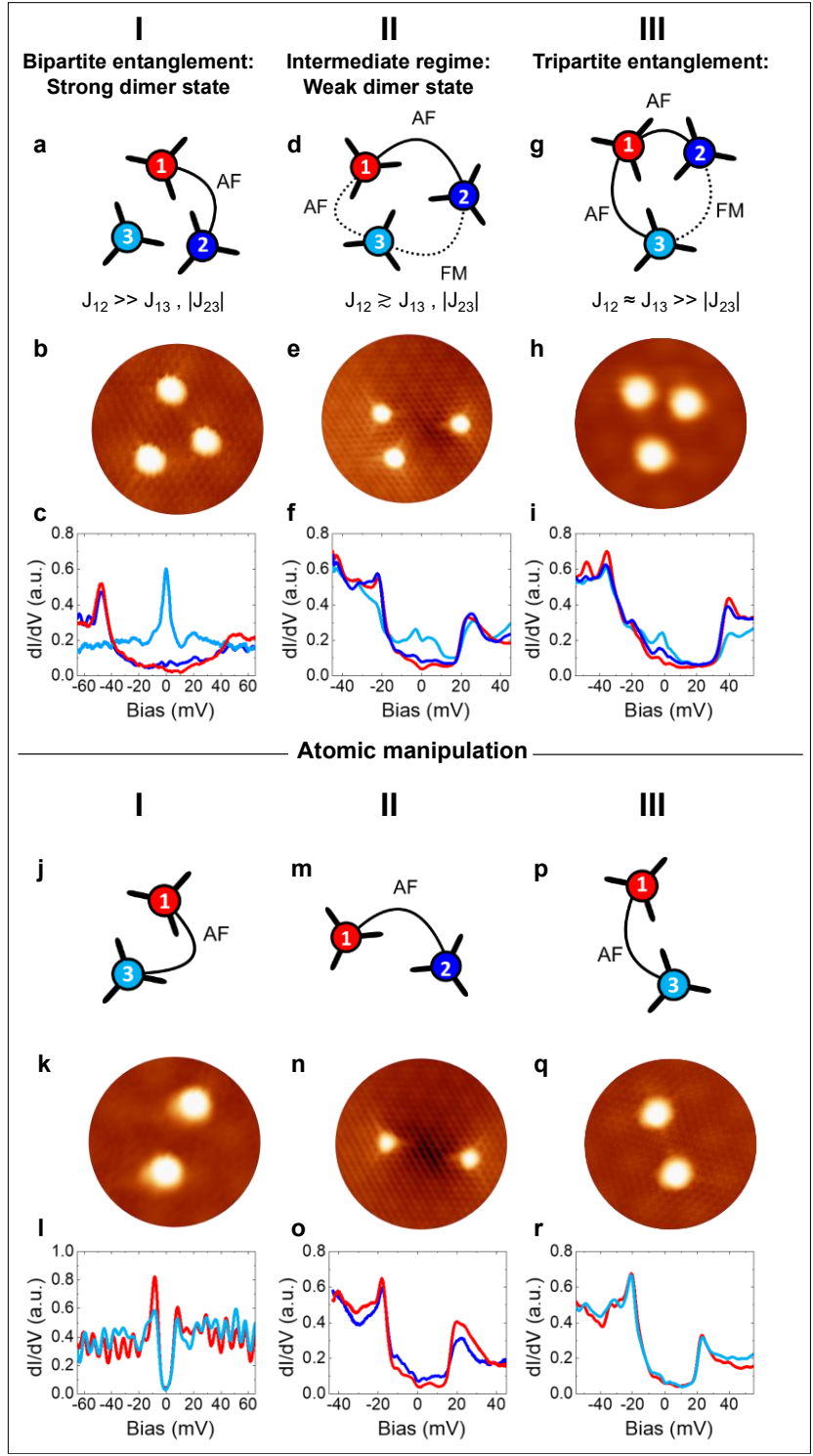

**Fig. 4. Three interacting S=1/2 spins: ABB S=1/2 configurations. a,** Schematics of bipartite entangled, dimerized ABB trimer. **b,** STM image of such ABB configuration (10mV, 0.1nA). **c,** *dI/dV* spectra measured on each of the H atoms in **b** (60mV, 0.1nA). **d,** Schematics of a weakly dimerized ABB trimer. **e,** STM image of such ABB configuration (4mV, 0.05nA). **f,** *dI/dV* spectra

measured on each of the H atoms in **e** (50mV, 0.1nA). **g,** Schematics of a tripartite entangled ABB trimer dominated by antiferromagnetic interactions. **h,** STM image of such ABB configuration (400mV, 0.05nA). **i,** *dI/dV* spectra measured on each of the H atoms in **h** (60mV, 0.1nA). **j,-r,** Corresponding schematics, STM images, and *dI/dV* spectra for each case (I-III) after atomic-scale manipulation.

In the first limiting case described earlier (weak coupling of the third spin), the strongly coupled antiferromagnetic dimer achieves maximal bipartite entanglement. As coupling to the third spin strengthens, entanglement monogamy dictates that bipartite entanglement within the dimer is progressively reduced, giving rise to genuine tripartite entanglement[42]. Due to the clear distinction in IETS signatures between these two extremes, we propose that spectral features observed in these systems could serve as measurable indicators of tripartite entanglement[43].

One could also expect that some trimers feature two clear inelastic steps, corresponding to transitions from the S=1/2 ground state to the excited states with either S=1/2 and S=3/2 (see SI1). However, in our experiments, we have not clearly identified that situation. We attribute this to the fact that the visibility of two steps is possible whenever the splitting of the two inelastic steps is larger than the IETS resolution (see Supplementary Fig. 1 and 6). In Supplementary Fig. 7 we plot a simulation of how the energy difference between excited states depends on the exchange couplings. We find that a significant fraction of the phase diagram does not comply with the visibility criteria.

In summary, our work demonstrates that the controlled hydrogenation of graphene down to the single-atom level unlocks robust, tunable magnetic moments with long-range and sublattice-dependent spin coupling. By assembling atomic-scale spin clusters—dimers, trimers, quartets, quintets, and hexamers (see Supplementary Fig. 8), and imaging their spin excitations with atomic precision, we establish a powerful platform for engineering designer spin Hamiltonians on demand. Looking ahead, combining this exquisite control with tunable carrier density promises to further tailor spin interactions in situ, while integrating superconducting proximity effects could open pathways to investigate exotic quantum phases at the confluence of magnetism and superconductivity[44,45]. Together, these advances herald new directions in quantum materials design and quantum simulation at the atomic scale.

## Methods

### Sample preparation and experimental details

Graphene was grown epitaxially by thermal decomposition of SiC[46,47], and atomic H was randomly deposited on the upmost graphene layer by cracking a molecular beam of $H_2$. Most chemisorbed H atoms (~95%) form non-magnetic dimers even for very dilute concentrations[48,49], thus, after H deposition, we use STM manipulation to build the desired magnetic H arrangements[28]. The magnetic configurations were fabricated using a two-step process: first H atoms were gathered with the STM tip by rapidly scanning the surface at low bias voltages (5–100 mV) and high tunneling currents (5–10 nA), second: the collected H atoms were then deposited onto a chosen graphene region using negative sample voltage pulses (up to −9 V). The measurements were performed on a low-temperature STM operating at 3 K. To facilitate spectroscopy measurements, we used Pb superconducting tips prepared, after H manipulation, by indentation on Pb nanoislands previously deposited on graphene. The presented *dI/dV* spectra were numerically deconvoluted to remove the effect of the superconducting tip unless indicated otherwise. By performing *dI/dV* curves on our configurations of H atoms, we can access the magnetic excitations of our system thanks to Inelastic Electron Tunneling Spectroscopy (IETS). The signature of spin excitations in IETS is an increase in the *dI/dV* spectra, symmetric at both bias polarities[19]. These excitation energies correspond to the exchange coupling between spins and thus provide a quantification of the strength of the magnetic coupling. The STM data was acquired and processed using the software WSxM[50].

**Theory methods**

We describe hydrogenated graphene using a Hubbard model in the mean field approximation,

$$H = t \sum_{\langle ij \rangle} \sum_{\sigma} a^{\dagger}_{i\sigma} a_{j\sigma} + U \sum_{i} \sum_{\sigma} \left[ \langle a^{\dagger}_{i\bar{\sigma}} a_{i\bar{\sigma}} \rangle a^{\dagger}_{i\sigma} a_{i\sigma} - \langle a^{\dagger}_{i\bar{\sigma}} a_{i\sigma} \rangle \langle a^{\dagger}_{i\sigma} a_{i\bar{\sigma}} \rangle \right],$$

where $t$ is the nearest-neighbor hopping integral, $U$ is the on-site (screened) Coulomb repulsion parameter, $\sigma=\uparrow,\downarrow$ denotes the spin direction, $i,j$ are atomic site indices and $a^{\dagger}_{i\sigma}$ ($a_{i\sigma}$) creates (annihilates) an electron state with spin $\sigma$ at site $i$. The angular brackets denote expected values in the mean-field configuration. As a chemisorbed hydrogen atom removes one p orbital and one electron from the honeycomb lattice, it can be modeled as a vacancy[51], keeping the ratio of electrons per carbon site equal to one (half-filling). We use a simulation cell with periodic boundary conditions with up to $N$=16000 sites. We adopt the value of 2.7 eV for the nearest-neighbor hopping integral $t$, and the Hubbard parameter $U=|t|$ for all the calculations shown. The excitation energy for a pair of chemisorbed H atoms is defined as the total energy difference between the ferromagnetic and antiferromagnetic mean-field configurations. Each magnetic configuration, characterized by the site dependent populations of the states associated with the two spin directions, is determined self-consistently. The populations for up and down spin states at each site are allowed to vary freely, under the following constraints: in the antiferromagnetic configuration the total number of electrons for both spin directions, $N_\uparrow$ and $N_\downarrow$, is the same ($N_\uparrow = N_\downarrow$); in the ferromagnetic configuration, $N_\uparrow = N_\downarrow+1$; for both configurations the total number of electrons is fixed at half-filling, $N_\uparrow + N_\downarrow = N$.

**Acknowledgements**

We acknowledge financial support from the Spanish Ministry of Science and Innovation, through project PID2023-149106NB-I00, the María de Maeztu Program for Units of Excellence in R&D (grant no. CEX2023–001316-M), the Comunidad de Madrid and the Spanish State through the Recovery, Transformation and Resilience Plan [Materiales Disruptivos Bidimensionales (2D), (MAD2DCM)-UAM Materiales Avanzados] the European Union through the Next Generation EU funds. B.V.-B. acknowledges funding from the Spanish Ministerio de Universidades through the PhD grant FPU2022/03675. J.F.-R., J.C.G.H. and A.C. acknowledge financial support from SNF Sinergia (Grant Pimag), FCT (Grant No. PTDC/FIS-MAC/2045/2021), and the European Union (Grant FUNLAYERS- 101079184). J.F.-R. acknowledges funding from Generalitat Valenciana (Prometeo2021/017 and MFA/2022/045) and MICIN-Spain (Grants No.PID2022-141712NB-C22 and PRTR-C17.I1). R.C. acknowledges funding from MICIU/EU/AEI through the FPI grant PRE2021-098139. E. C. R. acknowledges funding from the Alexander von Humboldt Foundation via the Henriette Herz program.

**Competing interests**

The authors declare no competing interests.

# Supplementary Information. Ultra-long-range spin coupling in graphene revealed by atomically resolved spin excitations

**Authors:** B. Viña-Bausá[1]*, A. T. Costa[2,3], J. Henriques[2,4], E. Cortés-del Río[5], R. Carrasco[1], P. Mallet[6], J-Y. Veuillen[6], J. Fernández-Rossier[2]*, I. Brihuega[1,7]*

**Affiliations:**

[1]*Departamento de Física de la Materia Condensada, Universidad Autónoma de Madrid; E-28049 Madrid, Spain*

[2]*International Iberian Nanotechnology Laboratory (INL); Avenida Mestre José Veiga, 4715-310 Braga, Portugal*

[3]*Physics Center of Minho and Porto Universities (CF-UM-UP), Universidade do Minho; Campus de Gualtar, 4710-057 Braga, Portugal*

[4]*Universidade de Santiago de Compostela, Santiago de Compostela, Spain.*

[5]*Department of Physics, University of Hamburg; D-20355 Hamburg, Germany.*

[6]*Université Grenoble Alpes, Grenoble, F-38400 France and CNRS, Institut Néel; Grenoble, F-38042 France*

[7]*Condensed Matter Physics Center (IFIMAC) and Instituto Nicolás Cabrera (INC), Universidad Autónoma de Madrid; E-28049 Madrid, Spain*

*\* Corresponding authors*

# Supplementary Information

1. Extensive study of magnetic coupling in AB and AA pairs

2. Spin excitation at higher temperatures

3. Distance dependence magnetic coupling and comparison with other systems

4. Mapping spin excitations and extension of coupled states

5. AAA configuration (S=3/2 ground state) and manipulation to ferromagnetic dimer

6. Heisenberg description of trimers

7. Analytical solution of Heisenberg trimer model

8. Building coupled structures with more than 3H atoms

## S1. Extensive study of magnetic coupling in AB and AA pairs

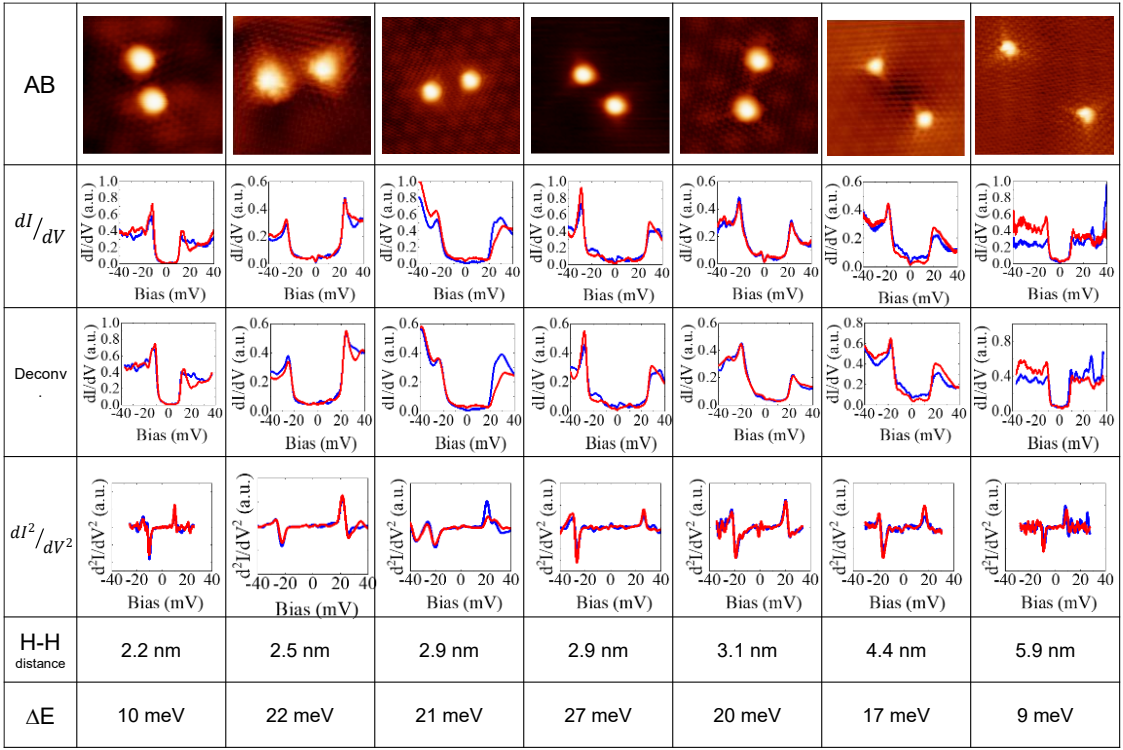

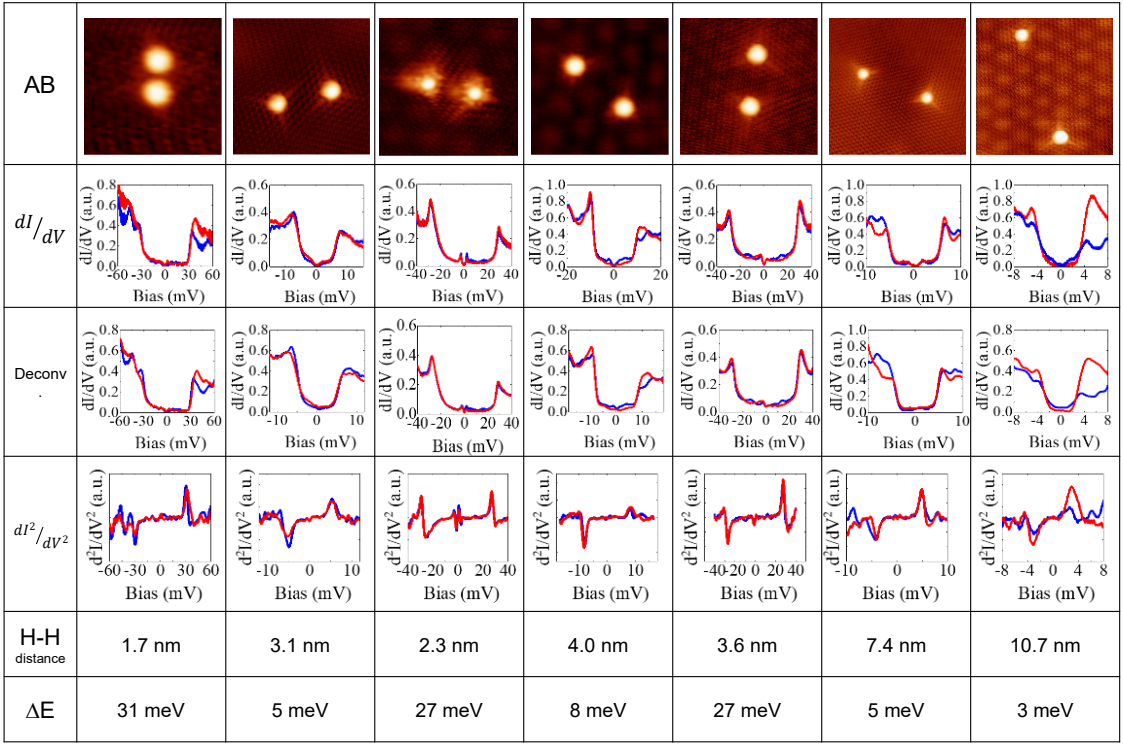

**Supplementary Table 1: Spin excitations in configurations of H atoms in opposite graphene sublattices.** In the first row we present the STM topography images of the configurations. In the second and third rows we show raw spectra measured with a superconducting tip and its numerical deconvolution respectively. The fourth row shows the numerical, second order derivative. The fifth and sixth row contain the distance between H atoms and the magnitude of the spin excitation, respectively.

| | | | | | | |
|---|---|---|---|---|---|---|
| AA | | | | | | |
| $dI/dV$ | | | | | | |
| Deconv. | | | | | | |
| $dI^2/dV^2$ | | | | | | |
| H-H distance | 0.5 nm | 1.9 nm | 2.4 nm | 2.7 nm | 3.6 nm | 9.0 nm |
| ΔE | 55 meV | 34 meV | 17 meV | 12 meV | 13 meV | 3 meV |

| | | | |
|---|---|---|---|
| AA | | | |
| $dI/dV$ | | | |
| Deconv. | | | |
| $dI^2/dV^2$ | | | |
| H-H distance | 4.7 nm | 2.4 nm | 2.2 nm |
| ΔE | 6 meV | 15 meV | 16 meV |

**Supplementary Table 2: Spin excitations in configurations of H atoms in the same graphene sublattices.** In the first row we present the STM topography images of the configurations. In the second and third rows we show raw spectra measured with a superconducting tip and its numerical deconvolution respectively. The fourth row shows the numerical, second order derivative. The fifth and sixth row contain the distance between H atoms and the magnitude of the spin excitation, respectively. Black lines have been included to identify the spin excitations in the second order derivative.

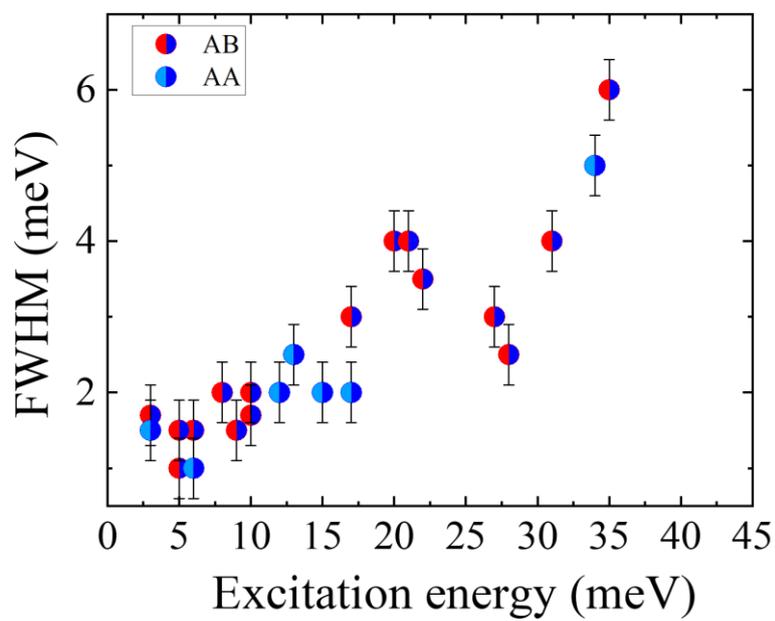

**Supplementary Fig. 1: IETS energy resolution.** Full width at half maximum (FWHM) of the inelastic spin excitations obtained from the numerical second derivative as a function of the excitation energy.

**S2. Spin excitations at higher temperatures**

AB pair from Chart S1 (bottom panel, column 3)

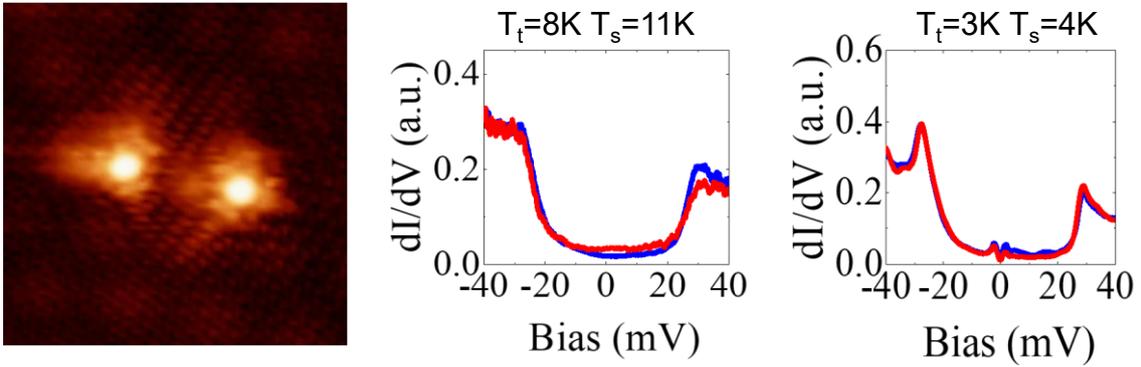

AB pair from Chart S1 (top panel, column 2)

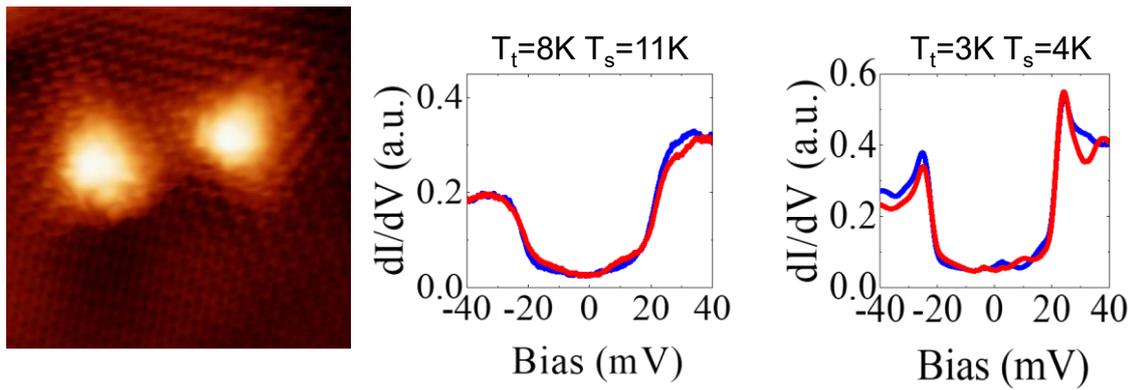

**Supplementary Fig. 2: Spin excitations in AB pairs measured with the sample at 11 K and the tip at 8 K.** (Top): AB from chart S1, corresponding to the configuration in the first column of the top panel. (Bottom): AB from chart S1, corresponding to the configuration in the third column of the top panel.

## S3. Distance dependence of magnetic coupling and comparison with other systems

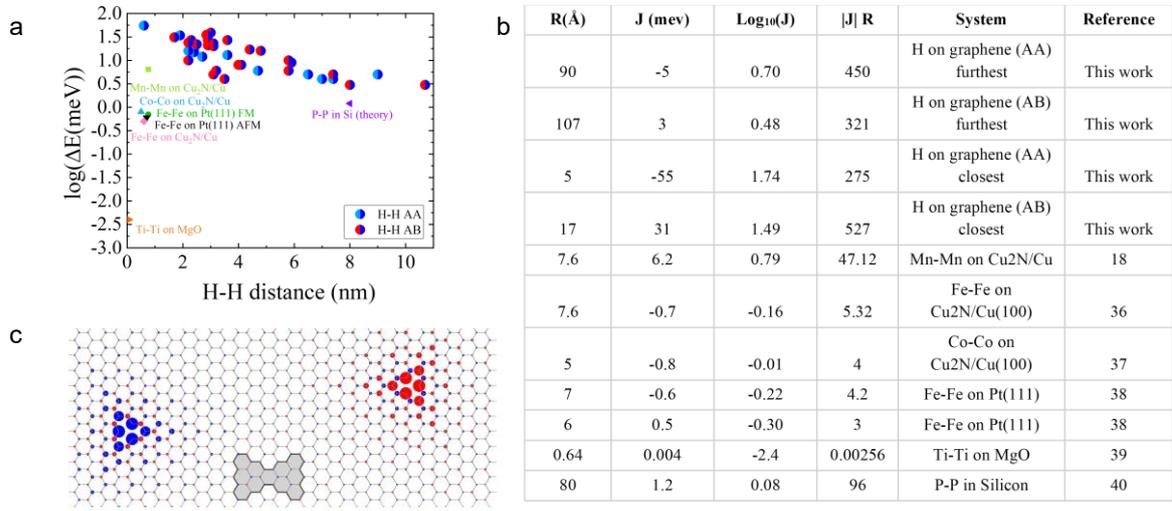

**Supplementary Fig. 3: Distance dependence of magnetic coupling and comparison with other systems. a,** Experimental exchange coupling in logarithmic scale as a function of distance between magnetic moments in hydrogenated graphene (this work) and in different compounds from refs. *18, 36-40*. **b,** Summary of the exchange energies and the interaction distance magnitude in this work and in different compounds from refs. *18, 36-40*. **c,** Comparison of lengths scales with nanographenes: Atomic magnetization map of the AB pair of H atoms of Fig1.c of the main text, with singlet-triplet splitting of 17 meV, superimposed with the molecular structure of Clar's goblet, with a measured[23] singlet-triplet splitting of 23 meV.

## S4. Mapping spin excitations and extension of coupled states

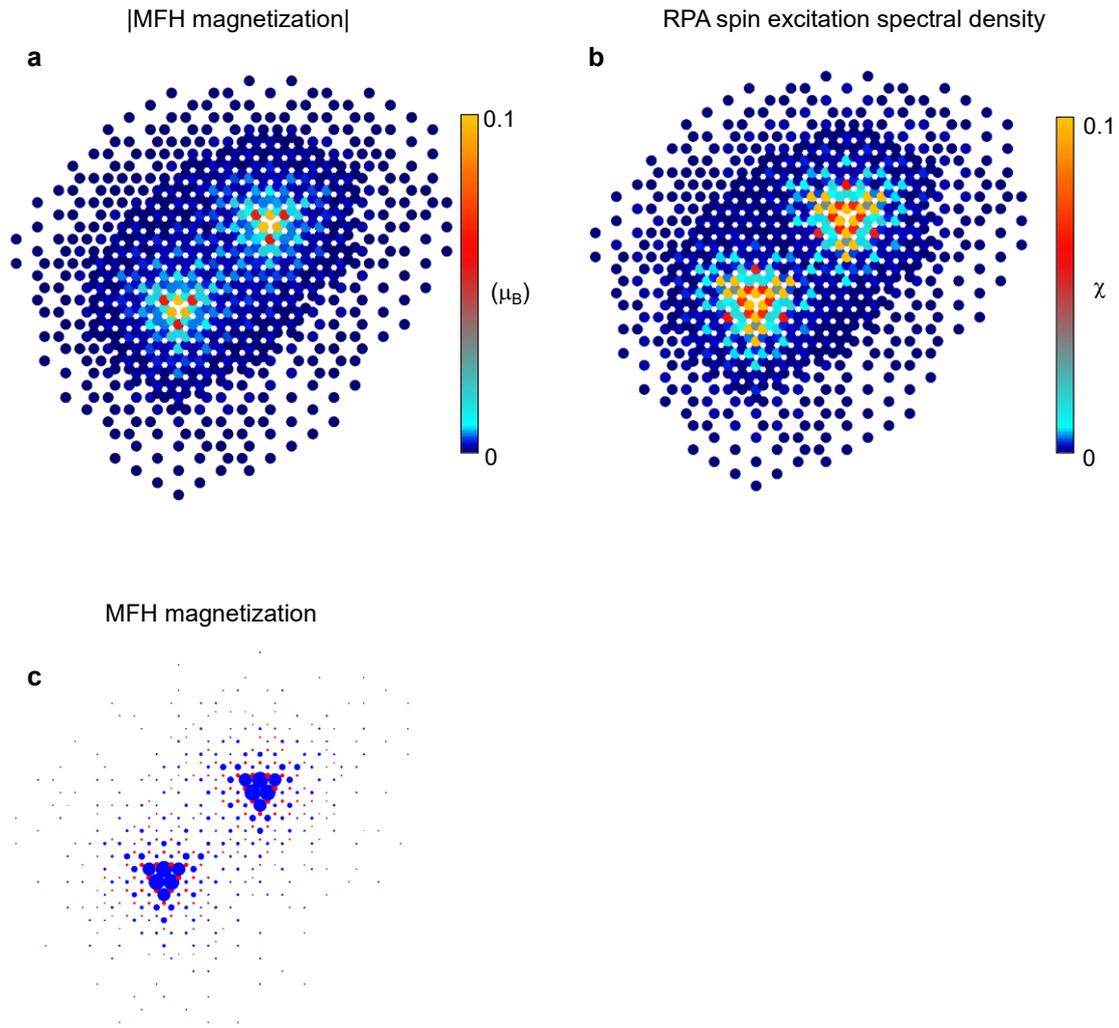

**Supplementary Fig, 4: AA MFH equilibrium magnetization map and random phase approximation (RPA) spectral weight of an AA configuration. a,** Absolute value of the MFH magnetization of an AA configuration at 2.4 nm. **b,** RPA calculated spectral density corresponding to the triplet to singlet excitation of the same configuration in **a**. The color scale has been adjusted to render a better comparison with the magnetization. **c,** Magnetization showing the ferromagnetic ground state of the configuration, dot size is proportional to the magnetization magnitude and blue and red indicate opposite spin orientations.

## S5. AAA configuration (S=3/2 ground state) and manipulation to ferromagnetic dimer

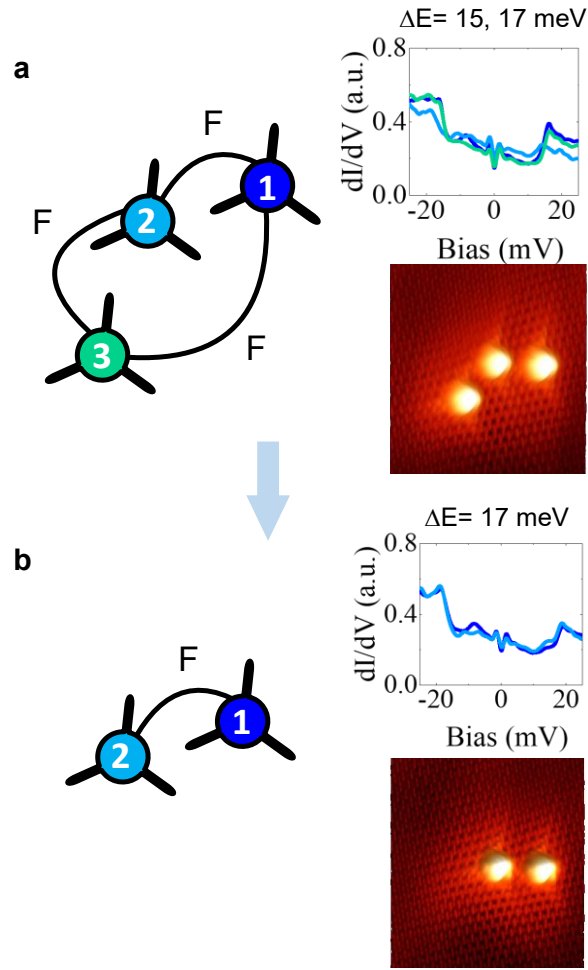

**Supplementary Fig. 5: AAA trimer. a,** *dI/dV* spectra, STM image and coupling diagram of an AAA trimer configuration. **b,** atomic scale manipulation of the configuration in **a**. *dI/dV* spectra, STM image and coupling diagram after the removal of H3.

## S6. Heisenberg model description of trimers

We solved the Heisenberg model using pairwise exchange couplings estimated from experiments as input parameters to model the trimers. Figure S1 presents the workflow and provides a comparative summary for the experimental configuration II shown in Fig. 4 of the main text. The calculated excitation energies from the Heisenberg model qualitatively agree with the experimental observations, highlighting its effectiveness in capturing key aspects of the system's magnetic behavior.

Nevertheless, the Heisenberg model does not fully describe the complete behavior observed in trimers. In particular, it does not reproduce the lower-bias features associated with the addition peaks originating from localized states of the hydrogen atoms (particularly pronounced at site 3). Additionally, the conductance step heights linked to the amplitude of spin excitations at each site are not precisely captured. Despite these quantitative differences, the relative increase in conductance at transition energies is qualitatively comparable between theory and experiment (see panel d).

The magnetic moments are delocalized over the graphene lattice, a factor expected to significantly shape the spatial distribution of excitations. Thus, reducing the system to a simplified three-site model may overlook important nuances. Importantly, the Heisenberg model identifies certain transitions close in energy, which might remain unresolved experimentally due to broader linewidths originating primarily from relaxation mechanisms involving graphene continuum states (see Extended Fig. 1).

Furthermore, the introduction of a third hydrogen atom could potentially introduce screening or enhancement effects in coupling, significantly influencing spin spatial distribution and magnetic transitions. These considerations motivate the development of more comprehensive models, capable of capturing beyond-pairwise exchange interactions, fermionic characteristics, and graphene-induced anisotropy. Exploring these aspects in greater detail will be the focus of future research.

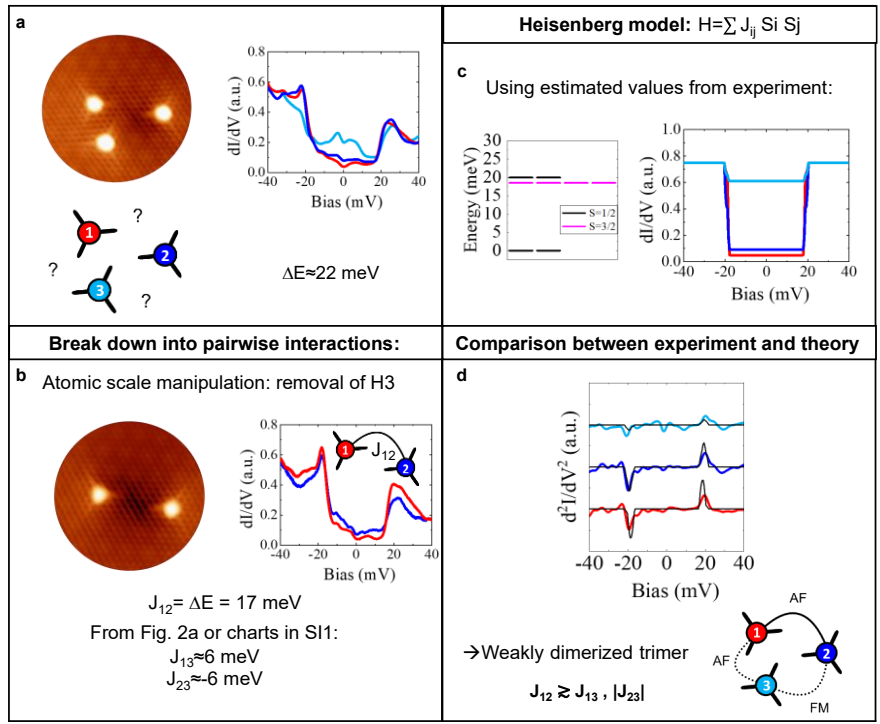

**Supplementary Fig. 6. Experimental measurements and modeling of trimers. a,** ABB trimer configuration with *dI/dV* spectra on the 3 sites showing a clear spin excitation. **b,** Atomic-scale

manipulation of the configuration in **a**. By removing H3 we can access the experimental exchange coupling between spins 1 and 2 ($J_{12}$). The remaining couplings ($J_{13}$, $J_{23}$) are estimated from the distance dependence plot and other experiments to try to fit best the measured excitation spectrum of the trimer. **c,** Simulated energy levels and *dI/dV* with Heisenberg model and tunneling current up to second order with the exchange couplings from **b**. **d,** Comparison of the experimental data (colors) and Heisenberg model (black lines) using the second derivative of the *dI/dV*.

**S7. Analytical solution of Heisenberg trimer model**

Let us consider the following Hamiltonian which describes three spins, interacting via linear exchange which is allowed to be different for each pair of spins:

$$H = J\left(\mathbf{S}_1 \cdot \mathbf{S}_2 + \alpha \mathbf{S}_2 \cdot \mathbf{S}_3 + \beta \mathbf{S}_1 \cdot \mathbf{S}_3\right)$$

where J > 0 is the exchange interaction between spins 1 and 2; Jα and Jβ are the exchange interactions between spins 2 and 3, and spins 1 and 3, respectively. Alternatively, we can write this Hamiltonian as

$$\frac{H}{J} = S_1^z S_2^z + \alpha S_2^z S_3^z + \beta S_1^z S_3^z$$
$$+ \frac{S_1^+ S_2^- + S_1^- S_2^+}{2} + \alpha \frac{S_2^+ S_3^- + S_2^- S_3^+}{2} + \beta \frac{S_1^+ S_3^- + S_1^- S_3^+}{2}$$

where we split the Hamiltonian into an Ising like, and a flip-flop contribution. The spectrum of this Hamiltonian is made of 8 states, which are arranged into two doublets (S = 1/2) and one quartet (S = 3/2). Since the Hamiltonian commutes with $S^2$ and $S^z$, we know that the states with $S^z = \pm 3/2$ are given by $|3/2, +3/2\rangle = |\uparrow\uparrow\uparrow\rangle$ and $|3/2, -3/2\rangle = |\downarrow\downarrow\downarrow\rangle$. These are eigenstates of the Hamiltonian, whose energy is entirely determined by the Ising part of the Hamiltonian:

$$H|3/2, +3/2\rangle = \frac{J}{4}(1 + \alpha + \beta)|3/2, +3/2\rangle$$
$$H|3/2, -3/2\rangle = \frac{J}{4}(1 + \alpha + \beta)|3/2, -3/2\rangle$$

To find the eigenergies of the = 1/2 states with $S^z = +1/2$ we note that these can be written in terms of the $S^2$ eigenstates:

$$|+\rangle = \frac{1}{\sqrt{3}}(|\downarrow\uparrow\uparrow\rangle + \omega_+|\uparrow\downarrow\uparrow\rangle + \omega_+^2|\uparrow\uparrow\downarrow\rangle)$$
$$|-\rangle = \frac{1}{\sqrt{3}}(|\downarrow\uparrow\uparrow\rangle + \omega_-|\uparrow\downarrow\uparrow\rangle + \omega_-^2|\uparrow\uparrow\downarrow\rangle).$$

Now, our goal is to express the Hamiltonian in this basis, and then diagonalize it. Computing the Hamiltonian matrix elements between the $|+\rangle$ and the $|-\rangle$ states, we find

$$\langle +|H|+\rangle = -\frac{1}{4}(1 + \alpha + \beta)$$
$$\langle -|H|+\rangle = \frac{1}{4}\left(-1 + i\sqrt{3} + 2\alpha - \left(1 + i\sqrt{3}\right)\beta\right)$$
$$\langle +|H|-\rangle = \frac{1}{4}\left(-1 - i\sqrt{3} + 2\alpha - \left(1 - i\sqrt{3}\right)\beta\right)$$
$$\langle -|H|-\rangle = -\frac{1}{4}(1 + \alpha + \beta)$$

Which gives the matrix representation

$$H = -\frac{J}{4}\begin{pmatrix} 1 + \alpha + \beta & 1 - i\sqrt{3} - 2\alpha + \left(1 + i\sqrt{3}\right)\beta \\ 1 + i\sqrt{3} - 2\alpha + \left(1 - i\sqrt{3}\right)\beta & 1 + \alpha + \beta \end{pmatrix}$$

whose eigenvalues read

$$E_\pm = -\frac{J}{4}\left(1 + \alpha + \beta \pm 2\sqrt{1 + \alpha^2 + \beta(\beta - 1) - \alpha(1 + \beta)}\right)$$

Thus, we find that the Heisenberg Hamiltonian for the trimer with inhomogeneous couplings gives three manifolds.

In the limit where one of the spins is weakly coupled to the other two, i.e. α, β ≪ 1, we find the following approximate expressions for the energy levels:

$$E(S = 3/2) = \frac{J}{4} + \frac{J}{4}(\alpha + \beta)$$
$$E(S = 1/2) \approx -\frac{3J}{4} + \mathcal{O}(\alpha\beta)$$
$$E(S = 1/2) \approx \frac{J}{4} - \frac{J}{2}(\alpha + \beta) + \mathcal{O}(\alpha\beta)$$

Thus, for the case where one of the spins is weakly coupled to the other one, the ground state is a S = 1/2 doublet, and the first excited state can be either S = 1/2 or S = 3/2 depending on α and β (which can be positive or negative). The absolute energy difference between the two excited manifolds, Δ, reads

$$\Delta = \frac{3J}{4}|(\alpha + \beta)|$$

Consider now the limit where two of the exchanges are similar, and have the same sign. Without loss of generality, we take α ≈ 1, and take β as the free parameter. Then:

$$E(S = 3/2) = \frac{J}{4}(2 + \beta)$$
$$E(S = 1/2) = -\frac{J}{4}\left(2 + \beta + 2\sqrt{(\beta - 1)^2}\right)$$
$$E(S = 1/2) = -\frac{J}{4}\left(2 + \beta - 2\sqrt{(\beta - 1)^2}\right)$$

Since, by definition, -1 < β < 1, then β - 1 < 0. Therefore:

$$E(S = 3/2) = \frac{J}{2} + \frac{J}{4}\beta$$
$$E(S = 1/2) = -J + \frac{J}{4}\beta$$
$$E(S = 1/2) = -\frac{3J}{4}\beta$$

where we once again find the ground state to be an S = 1/2 doublet. The first excited state is also a doublet, and the quartet appears as the highest energy state.

Alternatively, we now consider α ≈ −1, and take β as the free parameter. Then:

$$E(S = 3/2) = \frac{J}{4}\beta$$
$$E(S = 1/2) \approx -\frac{\sqrt{3}}{2}J - \frac{J}{4}\beta + \mathcal{O}(\beta^2)$$
$$E(S = 1/2) \approx +\frac{\sqrt{3}}{2}J - \frac{J}{4}\beta + \mathcal{O}(\beta^2)$$

As before we find the ground state to be an S = 1/2 doublet, only this time the first excited state is the S = 3/2 quartet, and the other S = 1/2 doublet appears as the highest energy manifold.

In Fig. S2 we plot the energy difference between the two excited states of a trimer as a function of the coupling parameters α and β and we locate the experimental configurations of Fig. 4 of the main text.

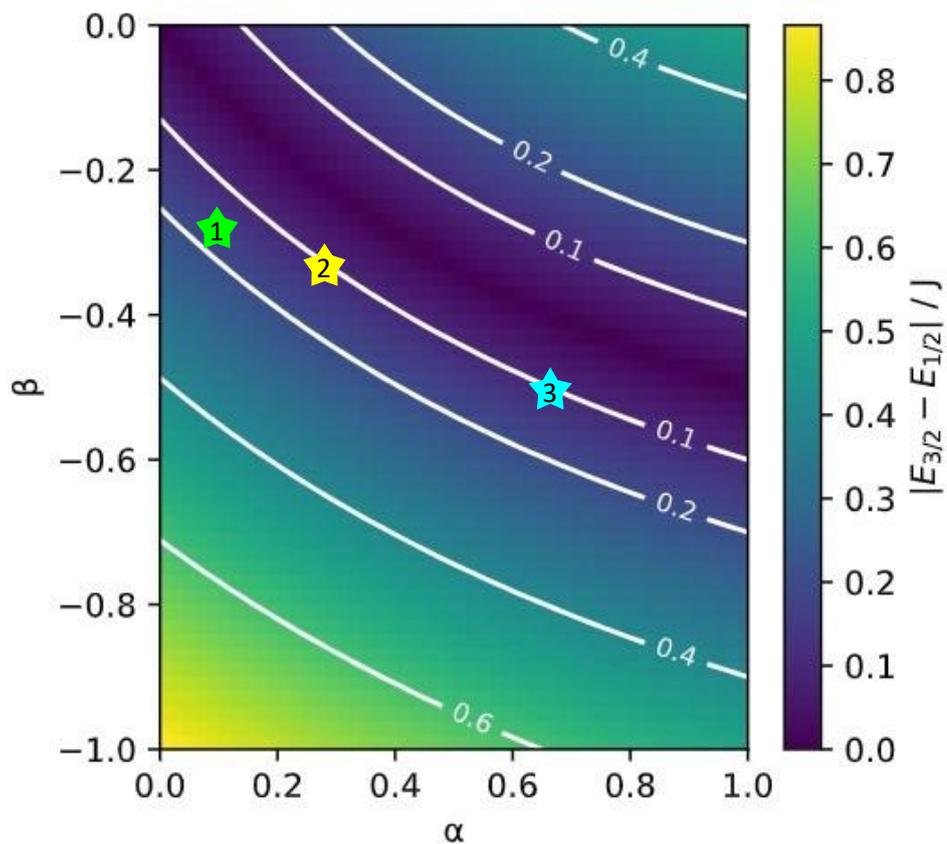

**Supplementary Fig. 7. Excitation spectra for ABB trimers as a function of the ratios of pairwise exchange couplings α and β.** Splitting between the trimer excited states for fixed J (J12) and for J23= β*J and J13= α*J. Stars correspond to the experiments I, II and III shown in Fig. 4 of the main text.

## S8. Building coupled structures with more than 3H atoms

In Fig. S3, we present an experiment beginning with a configuration of 8H atoms. The data reveal that the induced magnetic moments are collectively coupled, as evidenced by the appearance of multiple inelastic excitation features (highlighted by dashed lines in Fig. S3b). Further support for this coupling comes from sequential removal of individual H atoms, which leads to significant changes in the spectra (see Fig. S3c–h).

By examining the evolution *dI/dV* curves during each atomic-scale manipulation, we are able to track site-specific changes. For instance, on site 1 (H1), one prominent spin excitation shifts from nearly 30 meV to 12 meV. Across all manipulation steps, multiple spin excitations are consistently observed, marked with tentative dashed lines in the bias-symmetric features in the spectra and summarized in Fig. S3j.

These observations demonstrate the tunability of this spin system, highlighting its potential for scalable architectures based on engineered magnetic interactions at the atomic level.

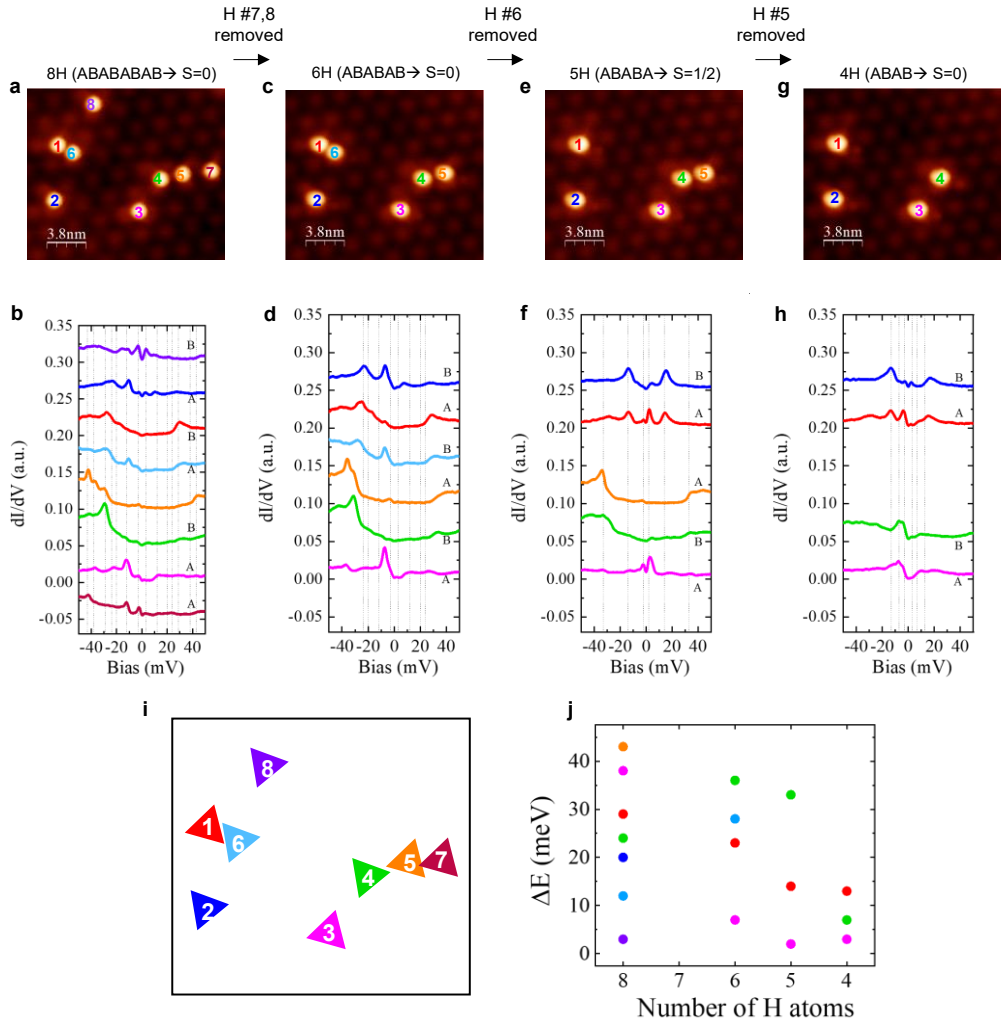

**Supplementary Fig. 8. Building coupled structures with more than 3H atoms, visualization of spin excitations and manipulation. a,** Configuration with 8H atoms and S=0. **b,** Spin excitations probed on each H site. Up to 7 spin excitations can be identified. **c,** Configuration with 6H atoms and S=0. **d,** Spin excitations probed on each H site. Up to 4 spin excitations can be identified. **e,** Configuration with 5H atoms and S=1/2. **f,** Spin excitations probed on each H site. Up to 3 spin excitations can be identified. **g,** Configuration with 4H atoms and S=0. **h,** Spin excitations probed on each H site. Up to 3 spin excitations can be identified. **i,** Schematics of the initial configuration with 8H atoms. **j,** Collective spin excitations in each situation with a different total number of H atoms.